\documentclass[graybox]{svmult}

\usepackage{mathptmx}       % selects Times Roman as basic font
\usepackage{helvet}         % selects Helvetica as sans-serif font
\usepackage{courier}        % selects Courier as typewriter font
\usepackage{type1cm}        % activate if the above 3 fonts are
\usepackage{makeidx}         % allows index generation
\usepackage{graphicx}        % standard LaTeX graphics tool
\usepackage{multicol}        % used for the two-column index
\usepackage[bottom]{footmisc}% places footnotes at page bottom

\usepackage{amsmath,amssymb,amsfonts}
\usepackage{algorithmic}
\usepackage{comment}
\usepackage{nomencl}
\usepackage{enumitem}   

\makeindex             % used for the subject index
\begin{document}

\title*{The Degree-Dependent Threshold Model: Towards a Better Understanding of Opinion Dynamics on Online Social Networks}
\author{Ece {{\c{C}}}i{{\u{g}}}dem Mutlu and Ivan Garibay}
\institute{Ece {{\c{C}}}i{{\u{g}}}dem Mutlu \at Complex Adaptive System Laboratory, University of Central Florida, Florida, USA \email{ece.mutlu@ucf.edu}
\and Ivan Garibay \at Name, Complex Adaptive System Laboratory, University of Central Florida, Florida, USA \email{ivan.garibay@ucf.edu}}

\maketitle
\abstract{With the rapid growth of online social media, people become increasingly overwhelmed by the volume and the content of the information present in the environment. The fact that people express their opinions and feelings through social media channels, influence other people and get influenced by them has led the researchers from various disciplines to focus on understanding the mechanism of information and emotion contagion. Threshold model is currently one of the most common methods to capture the effect of people on others' opinion and emotion. Although many studies employ and try to improve upon the threshold model, the search for an appropriate threshold function for defining human behavior is an essential and yet an unattained quest. The definition of heterogeneity in thresholds of individuals is oftentimes poorly defined, which leads to the rather simplistic use of uniform and binary functions, albeit they are far from representing the reality. In this study, we use Twitter data of size 30,704,025 tweets to mimic the adoption of a new opinion. Our results show that the threshold is not only correlated with out-degree of nodes, which contradicts other studies, but also correlated with nodes' in-degree. Therefore, we simulated two cases in which thresholds are out-degree and in-degree dependent, separately.  We concluded that the system is more likely to reach a consensus when thresholds are in-degree dependent; however, the time elapsed until all nodes fix their opinions is significantly higher in this case. Additionally, we did not observe a notable effect of mean-degree on either the average opinion or the fixation time of opinions for both cases, and increasing seed size has a negative effect on reaching a consensus. Although threshold heterogeneity has a slight influence on the average opinion, the positive effect of heterogeneity on reaching a consensus is more pronounced when thresholds are in-degree dependent.}

\section{Introduction}
\label{sec:1}
While studying networks is not new to humanity, its focus has evolved from physical proximity and socio-economic based to social media based networks. This change is arguably the product of the fast-paced information flow that is engendered by the technological advances of the 21st century, and the resulting impact on people's needs and lifestyles. The need to address the newly-emerged phenomenon that people create, receive and disseminate information on online social networks has amplified the interest in the field of network science. Indeed, network science applications have extended to the field of marketing \cite{iyengar2011opinion, culotta2016mining}, sociology \cite{lewis2018conversion}, political science \cite{shi2019wisdom}, physics \cite{ruan2015kinetics}, economics \cite{mishra2019evaluating}, and biology \cite{li2017scored}, in attempting to reveal the interdependency between units of interest. For instance, the shift from traditional advertising to digital marketing applications, or the political campaigns being organized on social media channels, has allowed people's opinions to be voiced freely and with far-reaching consequences. This reciprocity in information flow, the increase in the volume of information received and sent, and the ease of relaying information has made it imperative that researchers and practitioners understand the dynamics of information and opinion formation, propagation, and exchange \cite{watts2007influentials, vosoughi2018spread, ugander2012structural, bovet2019influence}.

In the mid-20th century, the field of sociology pioneered the development of information and opinion diffusion as a subject of study, which has remained relevant and popular to this day. One of the early studies is the Markovian linear threshold model introduced by Granovetter \cite{granovetter1978threshold}, which investigates the opinion dynamics of people. According to the threshold model, individuals adopt a new opinion only if a critical fraction of their neighbors have already adopted the new opinion. Granovetter suggests that the threshold of individuals can be different, and are influenced by demographic and psychographic factors such as socio-economic status, education, age, personality type, etc. However, this heterogeneity among researchers is poorly-defined, which leads to an extensive use of homogenous (uniform) \cite{liu2018impacts, sprague2017evidence,singh2013threshold} and binary \cite{wang2016dynamics} threshold in many studies. Arguably, this assumption of homogenous or binary thresholds is an oversimplification of reality and may produce misleading results. To remedy this oversimplification and thereby provide a more holistic and accurate model, more complex threshold models such as tent-like function \cite{zhu2018dynamics}, truncated normal distribution function \cite{karampourniotis2015impact} or sigmoid function \cite{fink2016investigating} are also used in the literature. Our Twitter data mining results show that threshold of an individual for adopting a new opinion (retweeting a tweet) is affected either by his out-degree (number of followers) or his in-degree (number of following/followee). Some studies have already employed degree-dependent threshold models in explaining the dynamics of information diffusion \cite{gleeson2013binary,lee2017social}, however the degree dependency of an individual's threshold is associated only with his out-degree. Additionally, these studies have implemented threshold heterogeneity by using custom threshold functions, which renders the results less robust and less reliable. Therefore, we want to analyze the sensitivity of information diffusion dynamics to in-degree and out-degree dependencies of thresholds. Another purpose of this study is to understand how threshold heterogeneity and network properties (seed size, mean-degree) affect information diffusion dynamics when thresholds are in-degree and out-degree dependent, separately.

The remaining part of this paper is structured as follows: First, we provide an overview of the Twitter data set and its subsequent analysis. Then, we describe the methodology that we used to generate networks, assign thresholds and run the simulations in the "Method" section. We give the results of simulations in the "Simulation Results" section. Finally, we discuss the results and explain the contributions of this study in the "Conclusion" section.

\section{Methods}
\subsection{Data set and Twitter Analysis Results}
The Twitter data set used for this study contains 30,704,025 tweets from the cybersecurity-related events from March 2016 to August 2017, of which 16,884,353 are retweets. We first collected follower and following counts of each user to relate the retweeting probability of users with the two aforementioned counts. We generated a matrix of which rows represent follower count clusters and columns represent following count clusters of all users in our data set (Figure 1.a). Then, we filtered users who have retweets only and generated the same matrix (Figure 1.b). Preliminary results show that majority of the users are clustered around the areas where follower and following counts are not extreme, and the matrix of retweeted users also show a similar pattern unsurprisingly. Since retweeting probabilities of users in each cluster, i.e. threshold for adopting a new opinion, are not clear from these matrices only; we calculated the element-wise division of these two matrices to figure out the ratio of number of retweeters to the number of all users in each cluster. Results show that the retweeting probability of users who have relatively lower following count is higher, i.e. threshold of a node seems to be positively correlated with his out-degree. On the other hand, the effect of varying follower count on the retweeting probability is not obvious since the left-bottom of the matrix is empty (Figure 1.c). Therefore, we extracted 3 most retweeted tweets (RT1, RT2, RT3) of retweet sizes 138,969, 58,546 and 57,280, respectively. We divided users into 8 clusters with respect to their follower (Figure 2.a-2.c) and following counts (Figure 2.d-2.f), independently rather than jointly clustering. For each clusters, we calculated the ratio of the number of users who retweeted RT1, RT2 or RT3 to number of all users, respectively, as in Figure 2.c. The only difference is that instead of all retweeters, we just focused on retweeters of RT1, RT2 and RT3. Thus, we could prevent the masking effect of non-active users in the whole data set. The results show that both follower and following count have a negative effect on the retweeting probability of users. Furthermore, we applied a one-sided Chi-square test ($\alpha=0.5$) to understand whether this decreasing pattern is statistically significant. We included relative $\chi^2$ values if the retweeting ratio in the cluster is significantly higher than that of next cluster (p-value is lower than 0.005). We observed that the retweeting probability decreases when follower count increases and this decreasing pattern is significant for almost all consecutive clusters. Nevertheless, the decrease between the consecutive clusters defined by following counts were significant only when following counts are not high. This is probably because clustering users according to their follower and following counts with the same limits affect the results of test statistics notably, since distributions of follower counts and following counts of users are not similar in the data set, i.e  the .8 and .9 quantiles and the maximum of the user following counts are 1916, 3860 and 3,136,215; while those are 2332, 5639 and 94,833,565 for user follower counts. When we decrease the number of clusters from 8 to 3 ($(0, 1K],(1K, 10K],(>10K]$), we observed that the retweeting probability decreases when following count increases and this decreasing pattern is significant for all consecutive clusters  ($\chi^2=\{12602.8, 18087,2 72.3\}$ for RT1, $\{2762.3, 1807.2, 52.0\}$ for RT2, $\{1299.3, 987.1, 87.1\}$ for RT3). Thus, our data analysis shows that thresholds of individuals to accept a new opinion is positively correlated with their in-degree and out-degree.

\begin{figure}[h!]
\sidecaption
\centerline{\includegraphics[width=1.0\textwidth]{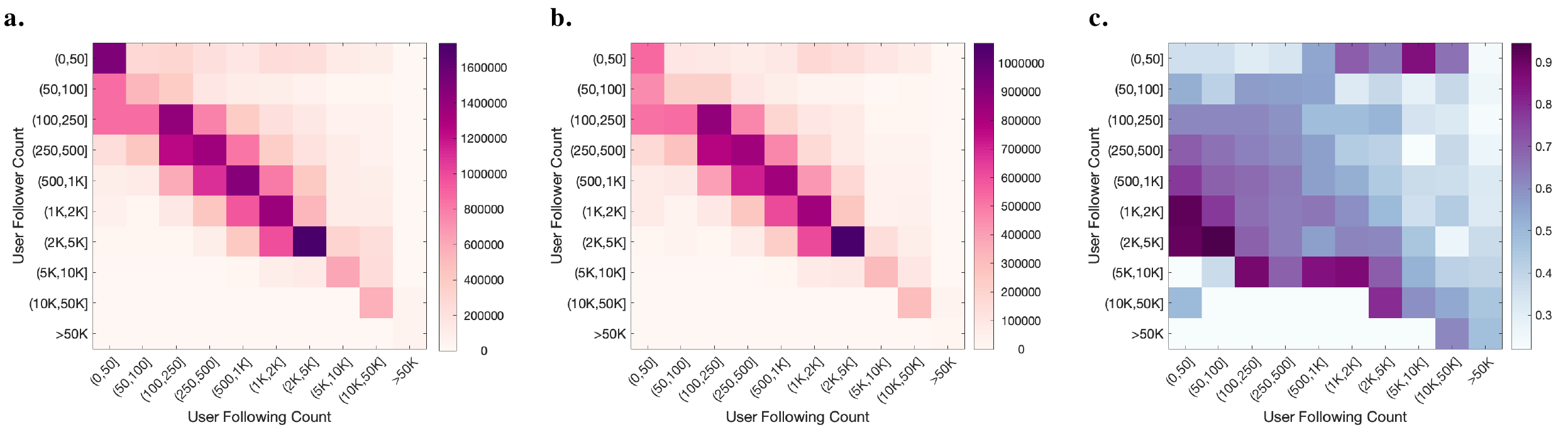}}
\caption{a. Number of users b. Number of retweeters c. The retweeting probability of users in each cluster -the element-wise ratio of number of retweeters in b to the number of users in a.}
\label{fig:1}       % Give a unique label
\end{figure}

\subsection{Generating Networks}
The main aim of this paper, as mentioned, is to investigate the effect of threshold heterogeneity on opinion spreading dynamics when thresholds are correlated with the degree-distribution of the nodes in a network. For this purpose, we generated power-law distributed random numbers ($x_i$) to further assign them to the desired degree-distribution of the network. To understand the effect of out-degree dependent threshold and in-degree dependent threshold on the dynamics of opinion spreading separately, we created two independent networks as:

\begin{enumerate}[label=(\roman*)]
  \item Out-degrees of the nodes ($k_{out}$) are power-law distributed and has the form $\sqrt{N}x^{\gamma}$ and in-degrees are kept constant ($M_{in}$).  
  \item In-degrees of the nodes ($k_{in}$) are power-law distributed and has the form $\sqrt{N}x^{\gamma}$ and out-degrees are kept constant ($M_{out}$).
\end{enumerate}

Here, $N$ denotes number of nodes in the network (seed size) and $\gamma=3$ for both cases for a fair comparison. Then, we added directed links between randomly selected node pairs ($i,j$) by employing configuration model \cite{newman2010networks} if $i\neq j$ and $k_{out} < x_i$ for i., $k_{in} < x_i$ for ii. This wiring process continued until all possible links are formed. In this network structure, self-edges are not allowed while multiple edges between same node pairs are possible. Since total in-degree in the network should be equal to the total out-degree in the network, one can easily realize that the mean-degree of the network is equal to:

\begin{enumerate}[label=(\roman*)]
  \item Fixed in-degree ($M_{in}$) when out-degrees of the nodes are power-law distributed. 
  \item Fixed out-degree ($M_{out}$) when in-degrees of the nodes are power-law distributed.
\end{enumerate}

\begin{figure*}[h!]
\centering
\centerline{\includegraphics[width=1.0\textwidth]{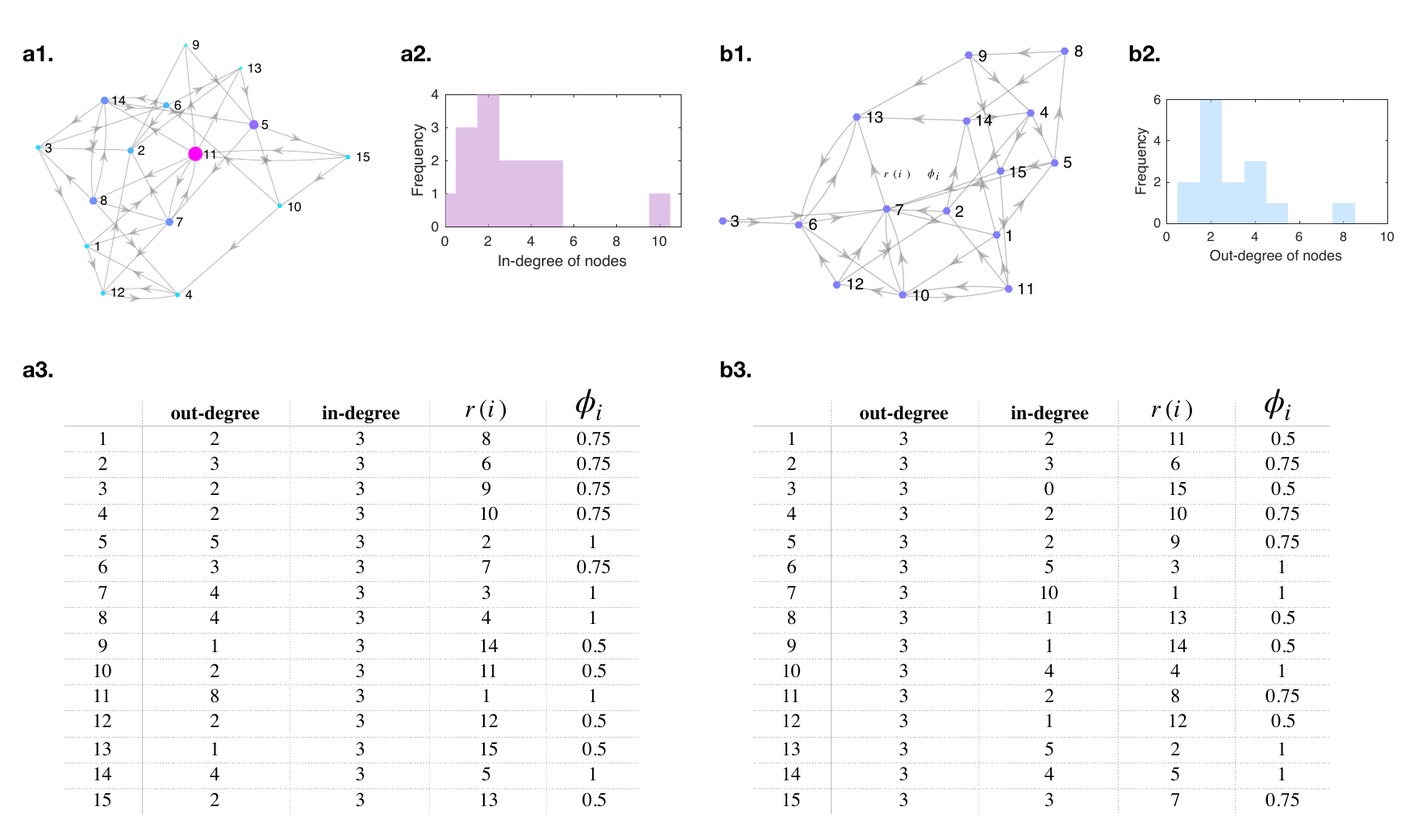}}
\caption{The representation of network when $N=15$ and a1. out-degrees are power-law distributed and in-degrees are kept constant as $M_{in}=3$, b1. in-degrees are power-law distributed and out-degrees are kept constant as $M_{out}=3$. Histogram plots of a2. out-degrees in a1, b2. in-degrees in b1. In addition to out-degree and in-degree of nodes, their ranks $r(i)$ and thresholds $\phi_i$ are also given in the table for a3. the network in a1, b3. the network in b1.}
\label{fig}
\end{figure*}

\begin{figure*}[t]
\centering
\centerline{\includegraphics[width=1\textwidth]{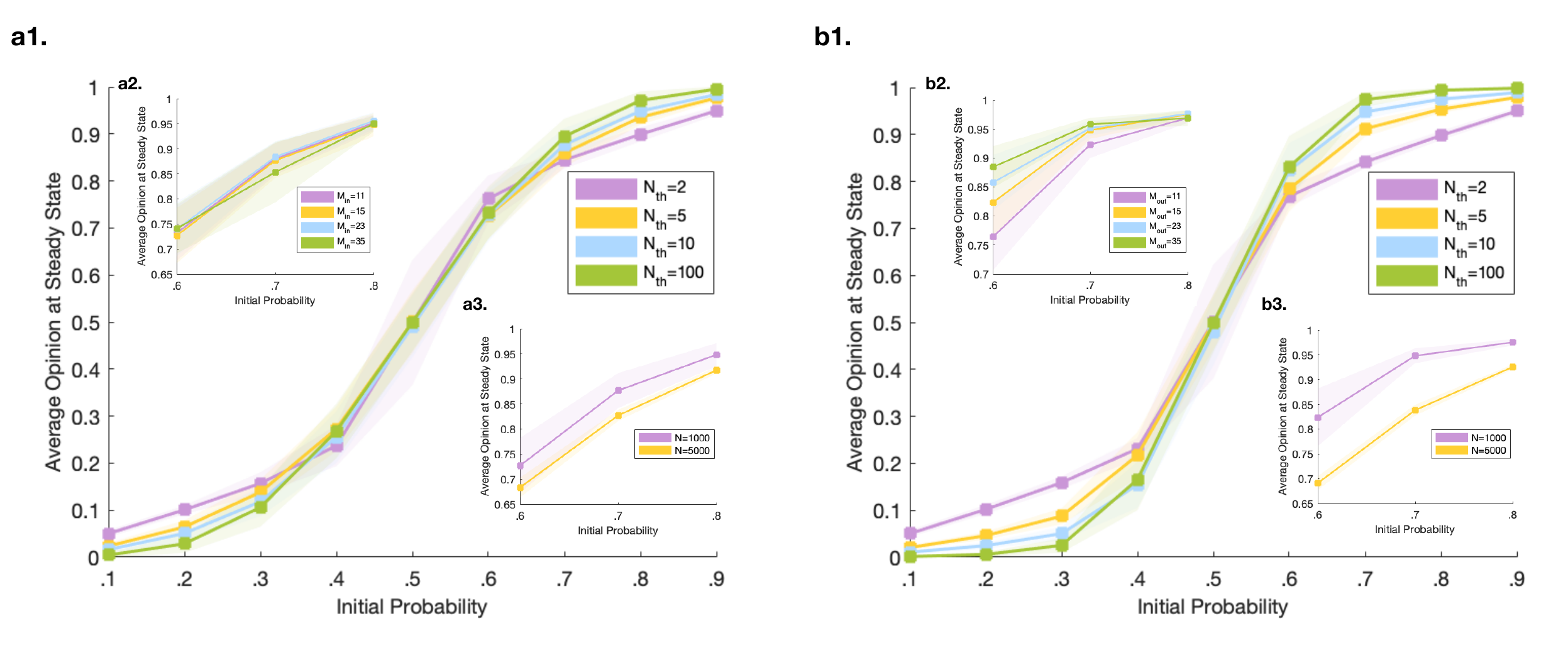}}
\caption{Simulation result of average opinion at steady state as a function of initial probability ($p$) a1. with varying threshold heterogeneity ($N_{th}$) when $N=1000$ and $M_{in}=15$, a2. with varying in-degree ($M_{in}$) when $N=1000$ and $N_{th}=10$ and a3. with varying seed size ($N$) when $M_{in}=15$ and $N_{th}=10$ if thresholds are out-degree dependent, and out-degrees are power-law distributed. Additionally, simulation result of average opinion at steady state as a function of initial probability ($p$) b1. with varying threshold heterogeneity ($N_{th}$) when $N=1000$ and $M_{in}=15$, b2. with varying out-degree ($M_{out}$) when $N=1000$ and $N_{th}=10$ and b3. as a function of initial probability ($p$) when $M_{out}=15$ and $N_{th}=10$ if thresholds are in-degree dependent, and in-degrees are power-law distributed.}
\label{fig}
\end{figure*}

\subsection{Assigning Thresholds}
After generating networks, we employed the degree-dependent threshold model by assigning the threshold of node $i$ to accept a new opinion ($\phi_i$) as correlated with:

\begin{enumerate}[label=(\roman*)]
  \item its out-degree when out-degrees are power-law distributed and in-degrees are constant in the network. 
  \item its in-degree when in-degrees are power-law distributed and out-degrees are constant in the network.
\end{enumerate}

Since threshold heterogeneity is one of our main concerns in this study, we divided nodes into $N_{th}$ groups by their ranks which can be obtained by sorting their
\begin{enumerate}[label=(\roman*)]
  \item out-degrees when out-degrees are power-law distributed and in-degrees are constant in the network.  
  \item in-degrees when in-degrees are power-law distributed and out-degrees are constant in the network.  
\end{enumerate}

Then, we assigned thresholds as evenly spaced $N_{th}$ points between 0.5 and 1 to prevent the confounding effect of the mean-threshold, i.e. the average threshold is always constant as 0.75. Thus, increasing $N_{th}$ yields more heterogeneity among thresholds of individuals.  

\[
  \phi_i =
  \begin{cases}
    0.5 & \text{if $r(i)\leq\frac{N}{N_{th}}$} \\
    0.5 + \frac{0.5}{N_{th}-1} & \text{if $\frac{N}{N_{th}}< r(i)\leq \frac{2N}{N_{th}}$} \\
    ... & ... \\
    0.5 + \frac{0.5 (N_{th}-2)}{N_{th}-1} & \text{if $\frac{(N_{th}-2)N}{N_{th}}< r(i)\leq \frac{(N_{th}-1)N}{N_{th}}$} \\
    1 & \text{if $\frac{(N_{th}-1)N}{N_{th}}< r(i)\leq N$} \\
  \end{cases}
\]
where $r(i)$ represents the rank of the node when they are sorted according to their i. out-degree and ii. in-degree. 

An example of network generation for two cases (i. and ii.), out-degrees and in-degrees and relative threshold values of the nodes are shown in Figure 3.

\subsection{Running Simulations}
We initialized the opinions of individuals as a Bernoulli distributed random variable with an initial probability ($p$), i.e. the opinion of the node $i$ ($s_i$) might equal to $1$ with a probability $p$ and equal to $0$ with a probability $1-p$. We assumed that the opinion change process is reversible; thus, individuals may change their opinions continuously rather than changing one time.  

After generating the network, assigning thresholds and initializing the opinions, we run the opinion change simulations. The process of updating their opinions is as follows:
\begin{enumerate}
    \item Picking a node $i$ randomly.
    \item Calculating the weighted average of the opinions of its in-neighbors ($\bar{o}_i$). Here, weights are the multiple edges formed between node $i$ and its neighbors. 
    \item Updating the opinion of node $i$ (${s}_i$) according to the criteria as follows:
    \begin{enumerate}
        \item \textbf{if} ${s}_i=0$ and $\bar{o}_i-{s}_i > \phi_i$, 
        \newline 
        \textbf{then} ${s}_i=1$ in the next step.
         \item \textbf{if} ${s}_i=1$ and $\bar{o}_i-{s}_i<-\phi_i$, 
         \newline 
         \textbf{then} ${s}_i=0$ in the next step.
    \end{enumerate}
\end{enumerate}

This Markovian chain is repeated until all possible opinion changes are made and individuals fix their opinion. We carried out all the simuluations on MATLAB and repeated these simulations for 10,000 times. 

\begin{figure*}[t]
\centering
\centerline{\includegraphics[width=1\textwidth]{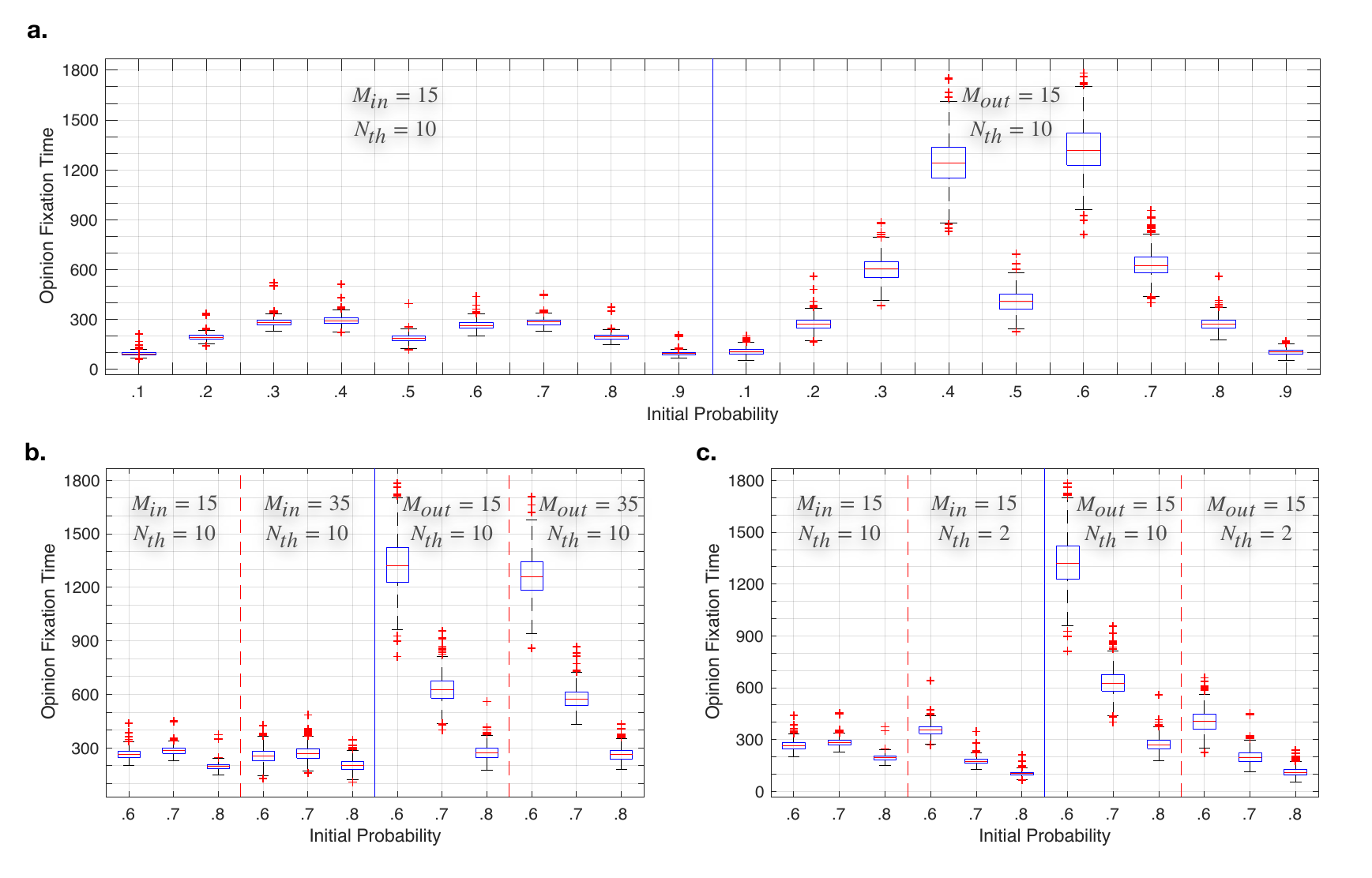}}
\caption{The comparison of fixation time of individuals as a function of initial probability a. when thresholds are out-degree dependent (left) and in-degree dependent (right) b. with varying mean-degree ($M_{in}$/$M_{out}$) when thresholds are out-degree dependent (left) and in-degree dependent (right) c. with varying threshold heterogeneity ($N_{th}$) when thresholds are out-degree dependent (left) and in-degree dependent (right).}
\label{fig}
\end{figure*}

\section{Simulation Results}
In the current study, we first aim to analyze the effect of in-degree and out-degree dependence of thresholds on the average opinion at steady state ($\bar{s}$). Therefore, after all the individuals fix their opinions in a network, we averaged the opinions of them by using the equation below:
\begin{equation}
    \bar{s}=\frac{1}{N}\sum_{i}^{N}{s_i}
\end{equation}
where $s_i$ is the opinion of node $i$ at steady state. We conducted our simulations to measure $\bar{s}$ as a function of initial probability ($p$) with varying mean-degree ($M_{in}/M_{out}$), seed size ($N$) and threshold heterogeneity $N_{th}$. Line plots in Figure 4 represent
the expected value of 10,000 Monte Carlo simulations, and
shaded areas with the same colors denote the relative one standard deviation from the expected value of these simulations. Here, Figure 4.a1-4.a3 show the simulation results when thresholds are out-degree dependent and out-degrees are power-law distributed while in-degrees are kept constant. Figure 4.b1-4.b3, on the other hand, indicates the simulation results when thresholds are in-degree dependent and in-degrees are power-law distributed while out-degrees are kept constant.

Figure 5, on the other hand, shows the time elapsed until all individuals fix their opinion ($t_f$) as a function of $p$ with varying $M_{in}/M_{out}$ and  varying $N_{th}$. We did not simulate the effect of varying $N$ on $t_f$ since it is obvious that increasing the seed size causes more deviation in the opinions and increases $t_f$. 

Figure 4.a1 and 4.b1 show $\bar{s}$ as a function of $p$ at various $N_{th}$ values. Since the standard deviation of the simulations are highest in the range $0.35 \lesssim p \lesssim 0.65$, we especially focus on the results when $p \lesssim 0.35$ and $p \gtrsim 0.65$. In general, the system is more likely to reach to a consensus when thresholds are in-degree dependent, and there is a clear asymmetry before and after $p=0.5$ in both cases. Therefore, we just focused on the region $0.6 \leq p \leq 0.8$ for further analyses. Although threshold heterogeneity of nodes in the system has a slight effect in the resulting average opinion when thresholds are out-degree dependent, we can say that probability that the system reaches a consensus increases as threshold heterogeneity increases; and this increase is more pronounced when thresholds are in-degree dependent. E.g. $\bar{s}=0.8412$ when $N_{th}=2$, while $\bar{s}=0.9742$ when $N_{th}=100$ at $p=0.7$ (Figure 4.b1). This can be explained as follows: When $N_{th}=2$, thresholds are distributed as $0.5$ or $1$ and a node which has opinion 0 can change its opinion from 0 to 1 when $8$ neighbors or all of his neighbors have opinion 1 if $N_{th}=2$, respectively. On the other hand, thresholds may take values of $0.500, 0.625, 0.750, 0.875$ or $1.000$ when $N_{th}=5$, and a node can change its opinion when $8$, $10$, $12$, $14$ or all of his neighbors have opinion 1 if $N_{th}=5$, respectively. When the initial probability is higher than $0.7$, one may expect that a node has more than $10.5$ ($M_{in} x p=15x0.7$) neighbors who have opinion 1 initially, and exceeding thresholds are easier when thresholds are not equal to 1. Therefore, the number of nodes who has opinion 1 is higher at the steady state when threshold heterogeneity is higher. When it comes to the effect of heterogeneity on the opinion fixation time, $t_f$ increases with increasing $N_{th}$ when thresholds are in-degree dependent. When thresholds are out-degree dependent, on the other hand, the effect of $N_{th}$ on $f_t$ is very minimal and the relation between $N_{th}$ and $t_f$ depends on $p$, e.g. increasing $N_{th}$ causes the people to fix their opinions more lately when $p\gtrsim0.7$, while the effect is opposite when $0.6\gtrsim p\gtrsim0.7$

Figure 4.a2 and 4.b2 show $\bar{s}$ as a function of $p$ at various $M_{in}$ and $M_{out}$  values. Results show that the change in the mean-degree has no prominent effect on the average opinion at the steady state when thresholds are out-degree dependent; however, increasing mean-degree seems to facilitate reaching a consensus when thresholds are in-degree dependent if $p\lesssim 0.7$. If $p\gtrsim 0.7$ in the same case, $\bar{s}$ values are very close to each other again. Since standard deviations of the results are high, we can conclude that mean degree does not affect average opinion at steady state either when thresholds are in-degree dependent or out-degree dependent. This is not surprising when we redefine the threshold model. The threshold model basically take the ratio of the node's threshold to the average opinion of his neighbors and the node changes his opinion if the ratio is higher than 1. Since the ratio does not change with changing mean-degree when the initial, $\bar{s}$ is not affected from $M_{in}$ and $M_{out}$. In fact, we would expect $t_f$ to increase because the number of links between nodes increased and this increase will cause more changes in ideas, the results show that the change in mean-degree has no effect on $t_f$ when thresholds are in-degree or out-degree dependent. 

Increasing node size in the network decreases $\bar{s}$ significantly when thresholds are in-degree dependent, whereas it has very little effect when thresholds are out-degree dependent. Since we analyze the effect of seed size on $\bar{s}$ when $0.6\leq p \leq 0.8$, it means that there is more diversity in the opinions when seed size is higher, e.g. $\bar{s}=0.9480$ when $N=1000$, while $\bar{s}=0.7155$ when $N=5000$ at $p=0.7$. It means that almost \%70 of the population has opinion 1 and \%30 has opinion 0 when $N=5000$. Low standard deviation in the Monte Carlo simulations also demonstrates the consistency of simulation results in every trial. As we mentioned before, we did not simulate a case in which $N=5000$ for the analysis of opinion fixation time since we expect the result is not novel and obvious. 

\section{Conclusion}
People make decisions in their daily lives on shopping, career, politics and so on. Although it looks like we make these decisions by ourselves, other people have a great influence on us since we are not isolated from each other. While face-to-face interactions used to be a main communication tool in the past, today's communication happens mostly on social media. Therefore, social network analysis has become very important to understand the dynamics of opinion formation, change, and propagation. One of the most common methods used to understand these dynamics is the threshold model, in which individuals adopt a new  opinion only if a critical fraction of their neighbors have already adopted the new opinion. First studies of social contagion have used homogeneous binary threshold model due to its simplicity; however, people show different attitude to adopt a new opinion, which renders the use of heterogeneous thresholds a must. Even though more complex thresholds are used in social contagion analysis nowadays, none of them validates their model with real data analysis. The main novelty of this study is that the degree-dependency of thresholds is inferred by using real world Twitter data. Social data analysis show that the threshold of a node does not only depend on his out-degree but also depend on his in-degree. Although the examples of out-degree dependent threshold models can be found in some studies, we also examined the results of opinion change simulations either for the in-degree dependent threshold model and out-degree dependent threshold model. Another contribution of this study is to investigate the effect of heterogeneity in thresholds on reaching a consensus for the first time. Our simulations demonstrated that the system is more likely to reach a  consensus when thresholds are in-degree dependent, rather than being out-degree dependent; however, people change their opinion more and fix their opinion more later in this case. The more heterogeneity in the thresholds is more likely to result in consensus but reaching a consensus takes more time, which is more significant when threshold are correlated with in-degree of nodes. Additionally, increasing seed size in the network makes the formation of consensus more difficult regardless of the dependence of threshold to the in-degree or out-degree. Another important point is that, as mean degree increases, diversity in opinions of individuals decreases when thresholds are in-degree dependent while it has no effect when thresholds are out-degree dependent. For future works, one may cluster Twitter users by using transfer entropy analysis to understand the dependence of users' threshold to their in-degree and out-degree. Thus, degree-dependence in threshold might be modeled more reasonably.  

\makenomenclature
 
\nomenclature{$r(i)$}{The rank of the node $i$ when it is sorted with decreasing degree}
\nomenclature{$\phi_i$}{The threshold of the node $i$}
\nomenclature{$M_{in}$}{The fixed in-degree value when in-degree is kept constant}
\nomenclature{$M_{out}$}{The fixed out-degree value when out-degree is kept constant}
\nomenclature{$\lambda$}{The parameter of power-law distribution}
\nomenclature{$N$}{seed size}
\nomenclature{$N_{th}$}{The threshold heterogeneity}
\nomenclature{$p$}{The initial probability}
\nomenclature{$s_{i}$}{The status/opinion of node $i$}
\nomenclature{$\bar{o}_{i}$}{The average opinion of neighbors of node $i$ }
\nomenclature{$\bar{s}$}{The average opinion at steady state}
\nomenclature{$k_{out}$}{Out degree distribution when out-degrees are power-law distributed.}
\nomenclature{$k_{in}$}{In degree distribution when in-degrees are power-law distributed.}
\nomenclature{$t_f$}{The time elapsed until all nodes fix their opinion, opinion fixation time.}

\printnomenclature

\section*{Acknowledgment}
This work was partially supported by grant FA8650-18-C-7823 from the Defense Advanced Research Projects Agency (DARPA).


\begin{thebibliography}{99.}%
\bibitem{bovet2019influence}
Alexandre Bovet and Hern{\'a}n~A Makse.
\newblock Influence of fake news in twitter during the 2016 us presidential
  election.
\newblock {\em Nature communications}, 10(1):7, 2019.

\bibitem{culotta2016mining}
Aron Culotta and Jennifer Cutler.
\newblock Mining brand perceptions from twitter social networks.
\newblock {\em Marketing science}, 35(3):343--362, 2016.

\bibitem{fink2016investigating}
Clay Fink, Aurora~C Schmidt, Vladimir Barash, John Kelly, Christopher Cameron,
  and Michael Macy.
\newblock Investigating the observability of complex contagion in empirical
  social networks.
\newblock In {\em Tenth International AAAI Conference on Web and Social Media},
  2016.

\bibitem{gleeson2013binary}
James~P Gleeson.
\newblock Binary-state dynamics on complex networks: Pair approximation and
  beyond.
\newblock {\em Physical Review X}, 3(2):021004, 2013.

\bibitem{granovetter1978threshold}
Mark Granovetter.
\newblock Threshold models of collective behavior.
\newblock {\em American journal of sociology}, 83(6):1420--1443, 1978.

\bibitem{iyengar2011opinion}
Raghuram Iyengar, Christophe Van~den Bulte, and Thomas~W Valente.
\newblock Opinion leadership and social contagion in new product diffusion.
\newblock {\em Marketing Science}, 30(2):195--212, 2011.

\bibitem{karampourniotis2015impact}
Panagiotis~D Karampourniotis, Sameet Sreenivasan, Boleslaw~K Szymanski, and
  Gyorgy Korniss.
\newblock The impact of heterogeneous thresholds on social contagion with
  multiple initiators.
\newblock {\em PloS one}, 10(11):e0143020, 2015.

\bibitem{lee2017social}
Eun Lee and Petter Holme.
\newblock Social contagion with degree-dependent thresholds.
\newblock {\em Physical Review E}, 96(1):012315, 2017.

\bibitem{lewis2018conversion}
Kevin Lewis and Jason Kaufman.
\newblock The conversion of cultural tastes into social network ties.
\newblock {\em American journal of sociology}, 123(6):1684--1742, 2018.

\bibitem{li2017scored}
Taibo Li, Rasmus Wernersson, Rasmus~B Hansen, Heiko Horn, Johnathan Mercer,
  Greg Slodkowicz, Christopher~T Workman, Olga Rigina, Kristoffer Rapacki,
  Hans~H St{\ae}rfeldt, et~al.
\newblock A scored human protein--protein interaction network to catalyze
  genomic interpretation.
\newblock {\em Nature methods}, 14(1):61, 2017.

\bibitem{liu2018impacts}
Quan-Hui Liu, Feng-Mao L{\"u}, Qian Zhang, Ming Tang, and Tao Zhou.
\newblock Impacts of opinion leaders on social contagions.
\newblock {\em Chaos: An Interdisciplinary Journal of Nonlinear Science},
  28(5):053103, 2018.

\bibitem{mishra2019evaluating}
Shraddha Mishra, Surya~Prakash Singh, John Johansen, Yang Cheng, and Sami
  Farooq.
\newblock Evaluating indicators for international manufacturing network under
  circular economy.
\newblock {\em Management Decision}, 57(4):811--839, 2019.

\bibitem{newman2010networks}
Mark Newman.
\newblock {\em Networks: an introduction}.
\newblock Oxford university press, 2010.

\bibitem{ruan2015kinetics}
Zhongyuan Ruan, Gerardo Iniguez, M{\'a}rton Karsai, and J{\'a}nos Kert{\'e}sz.
\newblock Kinetics of social contagion.
\newblock {\em Physical review letters}, 115(21):218702, 2015.

\bibitem{shi2019wisdom}
Feng Shi, Misha Teplitskiy, Eamon Duede, and James~A Evans.
\newblock The wisdom of polarized crowds.
\newblock {\em Nature Human Behaviour}, page~1, 2019.

\bibitem{singh2013threshold}
Pramesh Singh, Sameet Sreenivasan, Boleslaw~K Szymanski, and Gyorgy Korniss.
\newblock Threshold-limited spreading in social networks with multiple
  initiators.
\newblock {\em Scientific reports}, 3:2330, 2013.

\bibitem{sprague2017evidence}
Daniel~A Sprague and Thomas House.
\newblock Evidence for complex contagion models of social contagion from
  observational data.
\newblock {\em PloS one}, 12(7):e0180802, 2017.

\bibitem{ugander2012structural}
Johan Ugander, Lars Backstrom, Cameron Marlow, and Jon Kleinberg.
\newblock Structural diversity in social contagion.
\newblock {\em Proceedings of the National Academy of Sciences},
  109(16):5962--5966, 2012.

\bibitem{vosoughi2018spread}
Soroush Vosoughi, Deb Roy, and Sinan Aral.
\newblock The spread of true and false news online.
\newblock {\em Science}, 359(6380):1146--1151, 2018.

\bibitem{wang2016dynamics}
Wei Wang, Ming Tang, Panpan Shu, and Zhen Wang.
\newblock Dynamics of social contagions with heterogeneous adoption thresholds:
  crossover phenomena in phase transition.
\newblock {\em New Journal of Physics}, 18(1):013029, 2016.

\bibitem{watts2007influentials}
Duncan~J Watts and Peter~Sheridan Dodds.
\newblock Influentials, networks, and public opinion formation.
\newblock {\em Journal of consumer research}, 34(4):441--458, 2007.

\bibitem{zhu2018dynamics}
Xuzhen Zhu, Wei Wang, Shimin Cai, and H~Eugene Stanley.
\newblock Dynamics of social contagions with local trend imitation.
\newblock {\em Scientific reports}, 8(1):7335, 2018.
\end{thebibliography}
\end{document}